\documentclass{iopart}
\usepackage{graphicx}
\usepackage{iopams,mathrsfs}
\usepackage{subfig}
\usepackage[bookmarks=false]{hyperref}

\newcommand{\tl}{\theta_{(\ell)}}
\newcommand{\tn}{\theta_{(n)}}
\newcommand{\tq}{\tilde{q}}

\newtheorem{theorem}{Theorem}

\begin{document}

\title{Trapped and marginally trapped surfaces in Weyl-distorted Schwarzschild solutions}
\author{Terry Pilkington$^1$, Alexandre Melanson$^1$, Joseph Fitzgerald$^1$ and Ivan Booth$^2$}
\address{$^1$Department of Physics and Physical Oceanography \\ Memorial University of Newfoundland,  NL A1B 3X7,  Canada}
\address{$^2$Department of Mathematics and Statistics \\ Memorial University of Newfoundland,  NL A1C 5S7,  Canada}
\eads{ibooth@mun.ca, tpilkington@mun.ca}

\begin{abstract} 

To better understand the allowed range of black hole geometries, we study Weyl-distorted Schwarzschild solutions. They always 
contain trapped surfaces, a singularity and an isolated horizon and so should be understood to be (geometric) black holes. However 
we show that for large distortions the isolated horizon is neither a future outer trapping horizon (FOTH) nor even a marginally trapped surface:
slices of the horizon cannot be infinitesimally deformed into (outer) trapped surfaces. 
We consider the implications of this result for popular quasilocal definitions of black holes.


\end{abstract}
\vspace*{1em}



\section{Introduction}

There are at least three distinct ways to characterize a black hole. The first is based on causal structure and defines the 
black hole region to be the set of all points that cannot send signals to future infinity. 
This region is the \emph{causal black hole} and its boundary is an \emph{event horizon}\cite{Hawking:1973}. 
The second characterization is geometric, defining the black hole region to be the union of all trapped surfaces in a spacetime. 
This is the \emph{total trapped region} or \emph{prison}\footnote{A nomenclature suggested by Don Page (private communication).}
(alternatively if one works from outer trapped surfaces it is the \emph{outer prison}).
The third is also geometric and tries to characterize the boundary rather than the interior of a black hole. Black hole boundaries
are identified as (outer) marginally trapped surfaces. These boundaries are variously known as \emph{trapping} \cite{Hayward:1994}, 
\emph{isolated} or \emph{dynamical horizons} \cite{Ashtekar:2004a}. Each of these characterizations 
has its own strengths and weaknesses and though they are all equivalent for the Kerr-Newman family of solutions, for more general 
spacetimes they are generally not equivalent. 

The inequivalence of the definitions has been most intensively studied for the
Vaidya spacetimes \cite{Bendov:2007, Krishnan:2007, Senovilla:2008, SenovillaBeng:2008, SenovillaBeng:2009, SenovillaBeng:2010}. 
Standard theorems governing (outer) trapped surfaces in asymptotically flat spacetimes guarantee that both the 
prison and outer prison are contained within the causal black hole \cite{Hawking:1973}. 
However, in the aforementioned papers it was demonstrated that while the outer prison saturates that region for 
Vaidya solutions, the boundary of the prison \emph{is not} the event horizon. 
Similarly despite the original intent of characterizing dynamical black hole boundaries with trapping \cite{Hayward:1994} or dynamical horizons, these studies demonstrate that the boundary of the (outer) prison is not one of 
these surfaces. 
Instead it turns out that horizons of these types are generically intersected by trapped surfaces. Further, by general geometric arguments, 
dynamical marginally (outer) trapped tubes can be shown to be non-rigid and may be continuously deformed \cite{SenovillaBeng:2010, Ashtekar:2005, Booth:2007}.
This non-uniqueness can be concretely demonstrated in the Vaidya spacetimes by direct numerical calculations 
which locate a variety of (intersecting) dynamical horizons  \cite{Krishnan:2007}; this is an example of the well-known fact that apparent horizons are 
foliation dependent. 
 
Thus, at least for dynamical spacetimes, black hole characterization is complicated. In this paper we take a couple of steps backwards and 
instead consider the corresponding situation for static spacetimes. One might hope that things will be more straightforward in these cases. 
To test this hope we study a class of Weyl spacetimes \cite{ExactSol} that represent distortions of a Schwarzschild black hole (see, for example 
\cite{Israel:1964, Doroshkevich:1965, Mysak:1966, Geroch:1982, Breton:1997, Fairhurst:2001, Frolov:2007, Abdolrahimi:2009}). No matter the 
strength of the distortion, there is always an isolated horizon in these spacetimes which encloses trapped surfaces and a singularity.  
Thus they are good geometric black hole candidates though in 
their pure vacuum form these spacetimes are not asymptotically flat and so do not contain an event horizon or causal black hole.


The paper is organized in the following manner. Section \ref{horizons} reviews the mathematics of trapped and marginally 
trapped surfaces along with the definitions and properties of the related quasilocal horizons.
Section \ref{dbh} discusses the Weyl-distorted Schwarzschild solutions and confirms that they always contain singularities and trapped surfaces.
Section \ref{IHprop} then identifies the isolated horizon in those spacetimes and finds the conditions under which it becomes either a 
marginally trapped surface or future outer trapping horizon. The general solution 
is quite complicated so Section \ref{sec:foliation} begins by examining some specific examples of distortions and foliations to help build our understanding. 
It then proceeds to examine quadrupole distortions in some detail and prove that for large distortions the isolated horizon is neither marginally 
trapped nor a future outer trapping horizon. Finally, Section \ref{Discussion} considers the implications of our result.

\section{Background}
\label{horizons}

\subsection{Trapped and marginally trapped surfaces}
Let $(M,g_{ab})$ be a four-dimensional time-orientable spacetime and let $S$ be a two-dimensional spacelike surface. We assume that there
is a good way to distinguish the inside and outside of $S$\footnote{Any notion of asymptotic infinity is sufficient as in that case the outside can be
deemed to be the side facing infinity. In our examples the outside will be that facing $r \rightarrow \infty$. }. Then we 
can assign $\ell$ and $n$ to respectively be future-oriented outward- and inward-pointing normal vectors. We cross-normalize them so 
that $\ell \cdot n = -1$. These conditions do not fully fix the scaling. Given any function $F$, if $\ell$ and $n$ are properly oriented and 
cross-normalized then the null vectors 
\begin{equation}
\ell \rightarrow \rme^F \ell \; \; \mbox{and} \; \;  n \rightarrow \rme^{-F} n
\label{rescale}
\end{equation}
also satisfy those conditions. The projection operator for this surface is 
\begin{equation}
\tq_a^{\; b} = g_a^{\; b} + \ell_a n^b + n^a \ell_b 
\end{equation}
and $\tq^{ab}$ is also the push-forward of the induced two-metric on $S$. 
%

For these normals, the outgoing and ingoing \emph{null expansions} are 
\begin{equation}
\theta_{(\ell)} = \tilde{q}^{ab} \nabla_a \ell_b \; \; \mbox{and} \; \; \theta_{(n)} = \tilde{q}^{ab} \nabla_a n_b \,  .
\end{equation}
Equivalently we may write these in terms of the variation of the two-metric under deformations:
\begin{equation}
\delta_\ell \tq_{ab} = \frac{1}{2} \tl \tq_{ab} + \sigma^{(\ell)}_{ab} 
  \; \; \mbox{and} 
  \; \; \delta_n \tq_{ab} = \frac{1}{2} \tn \tq_{ab} + \sigma^{(n)}_{ab}  \, , \label{dLtq}
\label{deformation}
\end{equation}
where the trace-free parts of the deformation are the \emph{shears}: $\sigma^{(\ell)}_{ab}$ and $\sigma^{(n)}_{ab}$.
In general, $\delta_X \Psi$ for some vector field $X$ and geometric quantity $\Psi$ is calculated by: 1) infinitesimally deforming 
$S$ in the direction $X$, 2) calculating $\Psi$ on the deformed surface and then  3) calculating the rate of change of $\Psi$ 
with respect to this deformation. Such variations are standard in differential geometry however a discussion of them using the 
same language and notation as this paper may be found in \cite{Booth:2007}. 


For a sphere in Minkowski space, one would expect $\tl > 0$ and $\tn < 0$: that is, area increases for outward deformations and decreases for inward
ones. Thus there is clearly something unusual happening if $\tl < 0$ and in this case we say that $S$ is \emph{outer trapped}. If both $\tl < 0$ and
$\tn < 0$ then we say that $S$ is \emph{(fully) trapped} (the variation of area is negative in all normal directions). There are also borderline cases. 
$S$ is a \emph{marginally outer trapped surface (MOTS)}  if $\tl = 0$ and a \emph{marginally trapped surface (MTS)} if $\tl = 0$ and $\tn <0$. 
The signs of the expansions (and so the classification) are not affected by 
the scaling of the null vectors. Under a rescaling (\ref{rescale}), $\tl \rightarrow \rme^F \tl$ and $\tn \rightarrow \rme^{-F} \tn$. 

The full extrinsic geometry of a two-surface is fixed by the expansions $\tl$ and $\tn$ and shears $\sigma^{(\ell)}_{ab}$ and $\sigma^{(n)}_{ab}$
along with the curvature of the normal bundle
\begin{equation}
\Omega_{ab} = d_a \tilde{\omega}_b - d_b \tilde{\omega}_a \, , 
\end{equation}
where $d_a$ is the induced covariant derivative on $S$ and
\begin{equation}
\tilde{\omega}_a = - \tq_a^b n_c \nabla_b \ell^c 
\end{equation}
is the connection on the normal bundle. The connection is perhaps better known as the \emph{angular momentum one-form} 
because for an isolated horizon the angular momentum associated with a rotational Killing vector field $\phi^a$ is 
\begin{equation}
J_\phi = \frac{1}{8 \pi } \int_S \tilde{\epsilon} \left( \phi^a \tilde{\omega}_a \right) \, . \label{AngMom}
\end{equation}
It is not hard to see that if ${\Omega}_{ab} = 0$ then the angular momentum associated with any such $\phi^a$ vanishes: the horizon is 
non-rotating. Note too that under rescalings (\ref{rescale}) the connection transforms as
\begin{equation}
\tilde{\omega}_a \rightarrow \tilde{\omega}_a + d_a F 
\label{rescaleTom}
\end{equation}
while the curvature is invariant (as of course it should be). 

In classifying the surfaces, the variations of the expansions are also of interest (these are essentially second derivatives of the area-forms). 
For vacuum spacetimes and marginally outer trapped surfaces we find
\cite{Booth:2007}:
\begin{equation}
\delta_\ell \tl = - \sigma^{(\ell)}_{ab} \sigma^{ab}_{(\ell)} \label{dLtL}
\end{equation}
and 
\begin{equation}
\delta_n \tl = - \frac{ \tilde{R}}{2} - d_a \tilde{\omega}^a + \tilde{\omega}^a \tilde{\omega}_a  
\label{dNtL}
\end{equation}
where $\tilde{R}$ is the Ricci curvature of $S$. For a general vector 
\begin{equation}
X^a = A \ell^a - B n^a \,  
\end{equation}
we have
\begin{equation}
\delta_X \tl = - d^2 B + 2 \tilde{\omega}^a d_a B - B \delta_n \tl + A \delta_\ell \tl \, . 
\label{dXtL}
\end{equation}
 
The variations of $\tl$  are used to specify a particularly interesting class of marginally outer trapped surfaces. 
By Hayward's classification \cite{Hayward:1994} a MOTS 
is an \emph{outer trapping surface} if there exists a scaling of the null vectors such that $\delta_n \tl < 0$\footnote{Note that outer 
is being used in two different ways in these phrases. For a MOTS, the outer refers to the direction in which the surface is 
marginally trapped while for an outer trapping surface the outer is used in analogy with the outer versus inner horizon of a Kerr-Newman 
black hole.}. 
Similarly if one is considering a MOTS which is contained in a spacelike hypersurface $\Sigma$ (essentially an apparent horizon on $\Sigma$), 
then it is said to be \emph{strictly stably outermost} if $\delta_X \tl < 0$ for some inward-pointing spacelike $X$ which is normal $S$ 
but parallel to $\Sigma$ \cite{Andersson:2005}. 
In both cases the extra condition is imposed to ensure that the MOTS may be infinitesimally deformed into 
an outer trapped surface. That is, the MOTS is the limit of a sequence of outer trapped surfaces. Similarly one may consider 
marginally trapped surfaces that can be infinitesimally deformed into fully trapped surfaces. Such surfaces have 
$\tl = 0$, $\tn < 0$ and $\delta_n \tl < 0$ and by Hayward's classification are called \emph{future outer trapping surfaces}.

\subsection{Stationary black hole boundaries}

Turning to black hole boundaries, a three-surface $H$ foliated by two-surfaces $(S_\lambda, \tq_{ab})$ 
is a \emph{trapping horizon} if the $S_\lambda$ are all marginally outer trapped \cite{Hayward:1994}. In this paper 
we are only interested in stationary horizons in vacuum and so immediately specialize to the case where $H$ is null with normal 
$\ell_a$ (and tangent $\ell^a$): then $H$ is a \emph{non-expanding} horizon \cite{AFK:2000}. By (\ref{dLtq}) and (\ref{dLtL}) and scaling
$\ell^a = (d/d \lambda)^a$ we have 
\begin{equation}
\mathcal{L}_\ell \tq_{ab} = \delta_\ell \tq_{ab} = 0 \, ; 
\end{equation}  
that is, the induced metric on the leaves of the horizon is invariant in time.
More generally, it is not hard to show that if $\tl=0$ for one foliation $S_\lambda$ of a non-expanding horizon, 
then it also vanishes for any other foliation $S'_\lambda$. In fact,
the intrinsic metric $\tq_{ab}$ is the same for all possible foliations. 

If there is a scaling of the null vectors such that 
\begin{equation}
\left[ \mathcal{L}_\ell, D_a \right] = 0 \, ,
\label{IHcond} 
\end{equation}
where $D_a$ is the induced covariant derivative on $H$, then a non-expanding horizon becomes an isolated horizon \cite{AFK:2000}.
For such a scaling not only the intrinsic but also the extrinsic geometry of $H$ is invariant in time. 
For example
\begin{equation}
\delta_\ell \tn = 0 \; \; \mbox{and} \; \; \delta_\ell \tilde{\omega}_a = 0 \; . 
\label{IH_Con}
\end{equation}


In many cases,  the quickest way to show that a particular null surface is an isolated horizon is to instead show that it is a
\emph{Killing horizon}: a null surface $H$ that is everywhere tangent to a global Killing vector field $\xi^a$ that reduces to $\ell^a$ on $H$. 
It is not hard to show that all Killing horizons are also isolated horizons (\cite{AFK:2000} again). Though isolated horizons are mathematically 
more general than Killing horizons the best known examples, including the event horizons in the 
Kerr-Newman family of black holes,  are also Killing horizons.

Isolated horizons obey a zeroth law. For any scaling of $\ell$ satisfying (\ref{IHcond}), one can show that the surface gravity, defined 
as the proportionality factor $\kappa$ satisfying
\begin{equation}
 \ell^a \nabla_a \ell^b = \kappa \ell^b  \, , 
\end{equation}
is constant over the horizon. Note however that its exact value is not fixed. If $\ell \rightarrow c \ell$ for any positive constant $c$, then 
$\kappa \rightarrow c \kappa$. Thus without some method  to choose a preferred scaling, all we can say for sure is that $\kappa$ is either positive,
negative or vanishing. For our purposes this will be enough. 


The boundaries of stationary black holes are isolated horizons, however the existence of an isolated horizon is not, in itself, sufficient 
to imply the existence of a black hole; there are examples of spacetimes that are entirely foliated by isolated horizons, but which
certainly do not contain any black hole \cite{PLJ:2004}. Thus, extra conditions are necessary and the most obvious possibility is to
extrapolate from Kerr-Newman and only identify an isolated horizon as a black hole boundary if it bounds a total trapped 
(or at least an outer trapped) region. 

A quasilocal way to (attempt to) enforce this condition 
is to require that $H$ be foliated by either outer trapping surfaces (if we want it to bound 
a region of outer trapped surfaces) or future outer trapping surfaces (for a region of fully trapped surfaces). Respectively these 
are \emph{outer trapping horizons (OTHs)} and \emph{future outer trapping horizons (FOTHs)}. 
Note however that, while appealing, in general there is no guarantee that these conditions are either necessary or sufficient. 
This is because most trapped and 
marginally trapped surfaces are ``wild'' \cite{Eardley:1998}: even for spherically symmetric spacetimes 
most trapped surfaces are highly irregular and do not share the spacetime symmetries. 
Let us consider the (outer) prison in a little more detail to see how the problems arise. 
 
First, in the introduction we noted that for dynamical spacetimes the boundary of the prison is not a trapping horizon. That said, in Vaidya
a spherically symmetric trapping horizon certainly exists, is foliated by two-surfaces with both $\tn < 0$ and $\delta_n \tl < 0$
and is contained in the prison. Then it is clear that even though this surface satisfies our conditions and is the uniform limit of a 
sequence of fully trapped surfaces, there must be other trapped surfaces that intersect and cross it. Thus it is contained within, rather than 
bounding, the prison and -- at least for dynamical spacetimes -- the conditions are not sufficient to ensure the horizon is truly a boundary. 

The wildness of the trapped surfaces might also negate necessity. To see this, note that it is possible that each point $p$ on $S$ could individually
be the limit of a sequence of points $\{p_i\}$ on a distinct set of 
(outer) trapped surfaces $\{S_i\}$. Thus each point on $S$ would be arbitrarily close to points on a (outer) trapped surface, however there would 
be no uniform limiting sequence of such surfaces which works for all $p \in S$ simultaneously. Mathematically, such a situation would be much 
less convenient than the uniform case. However, as we shall see, that doesn't mean that it can't happen.

All of that said the $\tl = 0$, $\tn < 0$ and $\delta_n \tl < 0$ conditions precisely identify the outer black hole horizon for the Kerr-Newman family 
of spacetimes. Thus, even though they run into problems for dynamical spacetimes, it is worth investigating how they function for other
isolated horizons. 

\subsection{Isolated (F)OTHs}

It is encouraging to note that in contrast to dynamical horizons, isolated outer trapping horizons are rigid. To see this, note that
rigidity is determined by whether or not $\delta_X \tl = 0$ has multiple solutions. However, if $\delta_\ell \tl = 0$, (\ref{dXtL}) reduces to 
become
\begin{equation}
0 = - d^2 B + 2 \tilde{\omega}^a d_a B - B \delta_n \tl \, . 
\end{equation}
Now, if $\delta_n \tl < 0$ then a maximum principle argument shows that the only solution in this case is $B=0$ \cite{Booth:2007}. 
Thus at least one of the problems associated with dynamical horizons falls away. 

We now consider the work that must be done to decide whether or not a given isolated horizon is either marginally trapped or a full-fledged FOTH. 
This is significantly more involved than showing that a null surface is an isolated horizon. To show that a surface is an isolated horizon 
one only needs to work with its null normal/tangent $\ell$. The particular foliation of $H$ into spacelike two-surfaces is irrelevant. This is especially 
obvious for the subset of Killing horizons. The definition of a Killing horizon makes no mention of either the foliation or the other null normal $n$.

In contrast $\tn$ is only defined in the presence of a foliation. Given $\ell^a$ there are many possible auxillary null one-forms. 
What selects a particular $n_a$ is the foliation $S_\lambda$ and requirement that $n_a$ be null, normal to 
the $S_\lambda$ and in the cotangent space of the horizon. However if all we care about is whether or not $\tn < 0$ everywhere, then the 
particular scaling of the null vectors doesn't matter: we saw earlier that under the standard rescalings $\tn \rightarrow \rme^{-F} \tn$ (which won't affect its sign). 
Thus, demanding that an isolated horizon $H$ be marginally trapped is really a restriction that there exist a foliation of $H$ with particular 
properties. There are no restrictions on how the null normals are scaled. 

By contrast, requiring that $\delta_n \tl < 0$ everywhere turns out to depend on the scaling alone and is independent of the foliation. To see this,
return to (\ref{dNtL}):
\begin{equation*}
\delta_n \tl = - \frac{ \tilde{R}}{2} - d_a \tilde{\omega}^a + \tilde{\omega}^a \tilde{\omega}_a  \, . 
\end{equation*}
The Ricci scalar and induced covariant derivative $d_a$ are determined entirely by the induced two-metric on $S_\lambda$ and 
we have seen that metric is same for all possible foliations of an isolated horizon. Thus, the only possible foliation dependence might
dwell in the angular momentum one-form $\tilde{\omega}_a$. Here though the properties of isolated horizons again come into play: 
for an isolated horizon $\tilde{\omega}_a$ is fixed up to the addition of a total derivative irrespective of the foliation (see for example 
\cite{Ashtekar:2000b,Ashtekar:2001jb}). Equivalently, it can be shown that for an isolated horizon 
\begin{equation}
\delta_{A \ell}  \tilde{\omega}_a = d_a (A \kappa_{(\ell)} ) \, , 
\end{equation}
a generalization of the second equation in (\ref{IH_Con}). 
That is, under deformations of the foliation, the angular momentum one-form only changes by a total derivative. 
Such changes in $\tilde{\omega}_a$ are indistinguishable from those that arise from rescalings (\ref{rescaleTom}). 
Thus, allowing for both refoliations and rescalings is equivalent to just considering rescalings.
All possible forms of $\delta_n \tl$ are exhausted by considering a free function $F$ in 
\begin{equation}
\delta_{\bar{n}} \theta_{(\bar{\ell})} = - d^2\! F + 2 \tilde{\omega}^a d_a F + \tilde{q}^{ab} (d_a F)(d_b F) + \delta_{n} \theta_{(\ell)} \, .
\label{dntl2}
\end{equation}
If there exists an $F$ such that this expression is everywhere negative, then a given isolated horizon is an outer trapping horizon. However, if
it can be proved that no such $F$ exists, it isn't. 

Somewhat surprisingly, it turns out these two apparently very distinct problems are equivalent \cite{Booth:2008}. For an isolated horizon, 
it can be shown (see for example \cite{Ashtekar:2001jb,Booth:2007}) that 
\begin{equation}
 \delta_\ell \tn  = - \kappa_{(\ell)} \tn - \frac{\tilde{R}}{2} + d_a \tilde{\omega}^a + \tilde{\omega}_a \tilde{\omega}^a \, , 
\end{equation}
but of course this vanishes (\ref{IH_Con}). The reappearance of the Ricci scalar and $\tilde{\omega}_a$ terms from $\delta_n \tl$ is 
immediately striking, but note the change in sign on the divergence term. This can be handled. Via the Hodge decomposition theorem, 
it can be shown that there is always a scaling of the null vectors $(\ell_o, n_o)$ such that $d_a \tilde{\omega}_o^a = 0$ \cite{Ashtekar:2001jb}. 
Then assuming rotational symmetry and scalings $\ell = \rme^{F} \ell_o$ and $\bar{\ell}  = \rme^{-F} \ell_o$ (where $F$ respects the symmetries)
it is straightforward to show that 
\begin{equation}
- d_a \tilde{\omega}^a + \tilde{\omega}_a \tilde{\omega}^a = d_a \bar{\omega}^a + \bar{\omega}_a \bar{\omega}^a
\end{equation}
and so
\begin{equation}
\kappa_{(\ell)} \tn = \delta_{\bar{n}} \theta_{(\bar{\ell})} \, . 
\label{equivalence}
\end{equation}
Now, the surface gravity is constant and so if it is positive, $\tn < 0$ if and only if $\delta_{\bar{n}} \theta_{(\bar{\ell})} < 0$. Thus if we can 
find a foliation such that $\tn$ is everywhere negative we will know that there also necessarily exists a scaling of the null vectors so that 
$\delta_n \tl$ is everywhere negative. Rotationally symmetric isolated horizons are FOTHs if and only if they are OTHs.

\section{Distorted Schwarzschild spacetime}
\label{dbh}
The outer horizons of the family of Kerr-Newman (asymptotically flat, deSitter, or anti-deSitter) solutions are all isolated FOTHs. 
Thus, to examine the utility of these horizons we must consider more general spacetimes. An obvious source
of such examples are the Weyl solutions: it is well-known that all static solutions of the Einstein equations may be 
written in this form. In particular one can use these techniques to smoothly distort standard black hole spacetimes including 
Schwarzschild \cite{Israel:1964, Doroshkevich:1965, Mysak:1966, Geroch:1982,Frolov:2007, Abdolrahimi:2009}, Reissner-Nortstr\"{o}m 
\cite{Fairhurst:2001} and (the stationary but non-static) Kerr  \cite{Breton:1997}. In this paper we will focus on Schwarzschild though our discussion makes use of ideas
developed for the more general cases. 

\subsection{Weyl-distorted Schwarzschild spacetimes}

Weyl-distorted Schwarzschid spacetime takes the form 
\begin{equation}
\fl
ds^2 = -\rme^{2U}\left(1 - \frac{2m}{r}\right)dt^2 + \rme^{-2U + 2V}\left[\frac{dr^2}{1 - \frac{2m}{r}} + r^2\,d\theta^2\right] + \rme^{-2U}\,r^2\,\sin^2\!\theta\,d\phi^2 \, . 
\label{metric}
\end{equation}
where $U \equiv U(r, \, \theta)$ and $V \equiv V(r, \, \theta)$ are axisymmetric potentials. 
From a manipulation of the vacuum Einstein equations, it can be seen that $U$ is necessarily a solution of the 
Euclidean $\mathbb{R}^3$ Laplace equation (see, for example, the nice discussions in \cite{Geroch:1982, Fairhurst:2001, 
Frolov:2007}). Such solutions may be split into two classes: those that diverge at infinity and those that
diverge at the origin. On the gravity side, these correspond to distortions respectively induced by either a 
non-asymptotically flat infinity or a distorted singularity. Here we are interested in solutions that are induced by external 
sources and so discard solutions that diverge at the origin. Then one possible expansion of $U$\footnote{In this case derived
from the solution in prolate spheroidal coordinates given in \cite{Breton:1997} via the transformations $x=(r-m)/m$ and $y = \cos \theta$. } is given by
\begin{equation}
\label{eq:U}
U(r,\,\theta) =  \sum_{i = 1}^{\infty} \alpha_i \left( \frac{R}{m} \right)^i P_i 
\label{U}
\end{equation}
where the $\alpha_i$ are (constant) multipole coefficients, the $P_k$ are Legendre polynomials (of a moderately complicated argument):
\begin{equation}
P_k \equiv P_k\left(\frac{(r - m)\cos\theta}{R}\right) \, 
\end{equation}
and
\begin{equation}
R  \equiv \left[ \left(1 - \frac{2m}{r}\right)r^2 + m^2\cos^2\!\theta \right]^{1/2} \, . 
\end{equation}
At first glance there might appear to be problems with an ill-defined $R$  for certain values of $\theta$ if $r < 2m$. However, this 
apparent problem is just an artifact of how we have written the solution. In each case, multiplying the $(R/m)$ term by the
Legendre polynomial will both remove the $R$ from denominators and leave only $R^2$ terms in the numerator -- thus it won't matter if 
$\left(1 - \frac{2m}{r}\right)r^2 + m^2\cos^2\!\theta$ is negative or vanishing.

Though the distorting potential $U$ obeys linear equations and so different solutions may be superposed, things are more complicated for $V$. Given a choice of $U$, the remaining Einstein equations determine $V$. However 
those equations are not linear in $U$; thus the possible $V$ do not superpose (again see the discussions of 
\cite{Geroch:1982, Fairhurst:2001, Frolov:2007}). The corresponding $V$ are
\begin{eqnarray}
\label{eq:V}
\fl
V(r,\,\theta) = \sum_{i = 1}^{\infty} \sum_{j=1}^\infty  \alpha_i \alpha_j \left(\frac{ij}{i+j}\right)  \left( \frac{R}{m}\right)^{i+j} \left(P_i P_j - P_{i - 1} P_{j-1} \right)\\
\fl \qquad  \qquad  - \frac{1}{m} \sum_{i = 1}^{\infty}\alpha_i\,\sum_{j = 0}^{i - 1}  \Big[(-1)^{i + j} \left(r - m\left(1 - \cos\theta\right)\right) 
 + r - m\left(1 + \cos\theta\right) \Big] \left( \frac{R}{m} \right)^j P_j \, . \nonumber  
\end{eqnarray}
That said, as we shall shortly see, we will not actually need to use this form for $V$; for our purposes
a knowledge of $U$ will be sufficient.


Our interest is in the spacetime in the vicinity of the horizon so we now 
consider the apparent singularity at $r=2m$. Given the similarity between (\ref{metric}) and Schwarzschild, it will come as no surprise that
it is just a coordinate singularity. Via an Eddington-Finkelstein-like transformation of the time coordinate,
\begin{equation}
t \to v - f(r,\theta) \label{EFtrans}
\end{equation}
where
\begin{equation}
f(r,\,\theta) = \int_{r_0}^{r}\frac{\rme^{-2U(y,\,\theta) + V(y,\,\theta)}}{1 - \frac{2m}{y}}dy \, , 
\label{hchoice}
\end{equation}
the metric becomes
\begin{eqnarray}
\fl
ds^2  =  - \rme^{2U} \left( 1 - \frac{2m}{r} \right) dv^2 + 2 \rme^V dv dr + 2 f_\theta \, \rme^{2U}  \left( 1 - \frac{2m}{r} \right) dv d\theta  
-2 f_\theta \, \rme^V  d r d\theta \nonumber \\
 + \left[r^2 \rme^{-2U + 2 V} + f_\theta^2 \,  \rme^{2U}  \left( 1 - \frac{2m}{r} \right)  \right] d\theta^2 
 + r^2  \rme^{-2U} \sin^2 \! \theta \, d \phi^2 \, , 
 \label{EFmetric}
\end{eqnarray}
where $f_\theta = \partial f/ \partial \theta$. Note that this is not
a unique transformation -- for each value of $r_o$ there is a different transformation and so a different $v$ coordinate. In particular
these give rise to different $v= \mbox{constant}$ foliations of $r=2m$. 
That said, each of these transformations eliminates the apparent singularity in the $dr^2$ coefficient. 

The only potential remaining singularities (apart from $r=0$) are conical. 
By the vacuum Einstein equations (in this case $R_{r \theta} = 0$)
\begin{eqnarray}
\left. \frac{\partial V}{\partial \theta} \right|_{r=2m} & = & 2 \left. \frac{\partial U}{\partial \theta} \right|_{r=2m} \Longrightarrow V(2m,\theta) = 
2 U(2m,\theta) - 2 u_o
\label{C1}
\end{eqnarray}
for some constant $u_o$. Thus 
the induced metric on the $v=\mbox{constant}$ slices of the $r=2m$ hypersurface is
\begin{equation}
dS^2 = 4m^2 \rme^{-2U} \left( \rme^{4(U-u_o)} d \theta^2 +  \sin^2 \! \theta d \phi^2 \right) \, , 
\label{twometric}
\end{equation}
independent of the choice of $r_o$ (we'll return to this fact in section \ref{IHprop}). 
To avoid conical singularities on these two-surfaces we require that  
\begin{equation}
U =  u_o \Longleftrightarrow V=0\; \; \mbox{for} \; \; \theta = 0 , \pi \, .
\label{C2}
\end{equation}   
Assuming both (\ref{C1}) and (\ref{C2}) forces extra restrictions on the coefficients $\alpha_i$. 
Specifically
 \begin{equation}
\sum_{k = 1}^{\infty}\alpha_{2k - 1} = 0  \; \; \mbox{and} \; \;   \sum_{k = 1}^{\infty}\alpha_{2k} = u_o \; .
\label{eq:multipoles}
\end{equation}

For later use note that the remaining Einstein equations imply that:
\begin{eqnarray}
\left. \frac{\partial U}{\partial r} \right|_{r=2m} = - \left. \frac{1}{2m} \left(\frac{\partial U}{\partial \theta} \cot \theta + \frac{\partial^2 U}{\partial \theta^2} 
\right) \right|_{r=2m} \label{E2}
\end{eqnarray}
and 
\begin{eqnarray}
\left. \frac{\partial V}{\partial r} \right|_{r=2m} = - \left. \frac{1}{2m} \left(\frac{\partial U}{\partial \theta} \cot \theta - \frac{\partial^2 U}{\partial \theta^2} 
- \left( \frac{\partial U}{\partial \theta}\right)^2 \right) \right|_{r=2m} \;  .  \label{E3}
\end{eqnarray}
Further on the horizon
\begin{equation}
U(2m, \theta) = \sum_{i=0}^{\infty} \alpha_i \cos^i \theta \, .  
\label{HorRel}
\end{equation}
These last three results along with (\ref{C1}) greatly simplify the algebra in later calculations. In particular they 
completely eliminate any need to work with the rather formidable (\ref{eq:V}).

\subsection{Do they contain black holes?}

For $U=V=0$ these solutions reduce to Schwarzschild and so certainly contain a black hole, while in the next section we will demonstrate that
they always contain an isolated horizon at $r=2m$. However, we have already noted that isolated horizons are not by themselves sufficient 
to imply the existence of black holes. Thus before continuing it is worth spending a bit of time confirming that these solutions are indeed black holes. 

To begin, there are certainly difficulties if we require the classical causal definition of a black hole. These spacetimes are 
not asymptotically flat: for non-trivial choices of the distortion coefficients $\alpha_i$, $U$ diverges as $r \rightarrow \infty$. 
The distortion at the horizon may intuitively be viewed as reflecting the distortion ``at infinity'' and for such a distorted infinity, the 
causal definition cannot be applied. Via a second interpretation of these solutions we can (partially) rescue the causal definition. 
In this view the metric (\ref{metric}) is understood to only  
describe the geometry in some neighbourhood of $r=2m$. However, at a global level, one assumes that farther out there is a matter distribution 
causing the distortion and beyond that distribution the spacetime is asymptotically flat (see, for example, the discussion in \cite{Geroch:1982}). 
It is generally assumed that standard spacetime surgery techniques could be used to construct such a spacetime. 
In that case there would be an event horizon at $r=2m$. There is some question as to whether physically reasonable matter could 
generate arbitrary distortions. For example, for general Weyl solutions $u_o$ can take any value.  However if one regards the distortion as being
caused by external matter fields the strong energy condition requires that $u_0 \leq 0$ \cite{Geroch:1982}. 



Luckily we have good reason to interpret these solutions as black holes even without the use of spacetime surgery to insert matter. 
First the singularity at $r =0$ is real. Like the rest of the spacetime it is distorted, but it is certainly still there. This can be seen by 
checking the Kretschman scalar, however as the resulting expression is not particularly concise, it is probably easier to just read the
extensive discussion in \cite{Frolov:2007}. 

Second, these solutions contain fully trapped surfaces. It is well beyond the scope of this paper to identify the full extent of the prison. However it is 
straightforward to see that, close to the singularity, trapped surfaces exist. In this regime (away from the horizon)
it is most convenient to return to the original metric (\ref{metric}). Then for surfaces of constant
$t$ and $r$, a future-pointing unit normal timelike vector is:
\begin{equation}
\hat{u}^a = - \rme^{U-V} \sqrt{\frac{2m}{r}-1}  \left[ \frac{\partial}{\partial r} \right]^a
\end{equation}
while the orthogonal outward-pointing unit normal spacelike vector is:
\begin{equation}
\hat{r}^a = - \frac{\rme^{-U}}{\sqrt{\frac{2m}{r}-1}  }  \left[ \frac{\partial}{\partial t} \right]^a \, . 
\end{equation}
A straightforward calculation shows:
\begin{equation}
\theta_{(\hat{u})} = \tq^{a b} \nabla_a \hat{u}_b = -  \frac{\rme^{U-V}}{r} \sqrt{\frac{2m}{r}-1} \left(2 + r \frac{\partial V}{\partial r} - 2 r  \frac{\partial U}{\partial r} \right) 
\end{equation}
\begin{equation}
\theta_{(\hat{r})} =  \tq^{a b} \nabla_a \hat{r}_b = 0 \, , 
\end{equation}
and so for $\ell = \hat{u}+\hat{r}$ and and $n = (\hat{u}-\hat{r})/2$ we have 
\begin{equation}
\tl = 2 \tn = \theta_{(\hat{u})} \, . 
\end{equation}
The $\theta_{(\hat{r})}$ curvature vanishes because the surfaces of constant $t$ are extrinsically flat (just as they are outside the horizon). 

Now, for general $U$ and $V$ and $r<2m$, $\tl$ and $\tn$ do not have constant sign and are negative in some places but positive in others.
This is most easily seen by direct calculation; for example with a pure quadrupolar distortion with large $\alpha_2$. However close to the 
singularity both $\tl$ and $\tn$ become negative. $U$ and $V$ are everywhere well-defined and in particular neither they nor their derivatives 
diverge to infinity. Then as $r \rightarrow 0$: 
\begin{equation}
 \tl = 2 \tn \rightarrow -    \frac{ 2 \sqrt{2m}}{r^{3/2}} \rme^{U-V} \, . 
\end{equation} 
Thus, no matter how large the distortion, if we get close enough to $r=0$ both expansions become negative. 

We will learn more about near-horizon trapped surfaces in the next two sections as we study the isolated horizon.

\section{The isolated horizon and its properties}
\label{IHprop}

Given the similarity in structure between the Schwarzschild metric and  (\ref{metric}) one might expect an isolated horizon to exist at $r=2m$. 
This is indeed the case. Rather than checking the definition directly, in this case it is easier to show that $r=2m$ is a Killing horizon and then 
recall that all Killing horizons are automatically isolated horizons as well. 
From (\ref{EFmetric}): 1) $\xi = \partial/\partial v$ is a global Killing vector field, 2) $r=2m$ is 
a null surface and 3) $\xi$ is tangent to $r=2m$. Then by the discussion surrounding (\ref{IH_Con}), it is also an isolated horizon and it 
is not surprising that the induced metric (\ref{twometric}) on slices of the horizon is independent of the choice of foliation. 

For $\ell = \xi$ the  surface gravity is 
\begin{equation}
\kappa_{(\ell)} = \frac{1}{4m}\rme^{2u_0} > 0 \, .  \label{kappa}
\end{equation}
Thus if we find a foliation of these horizons that satisfies $\tn < 0$ it will also automatically satisfy $\delta_n \tl < 0$. We will explicitly explore
this equivalence in the following subsections. 

\subsection{An expression for $\tn$}

We begin by finding an equation for $\tn$. Since our distortions are rotationally symmetric, 
we restrict our attention to similarly rotationally symmetric foliations with leaves defined by
\begin{equation}\label{eq:surfaces}
S_\lambda = \bigg\{\left(v,\,2m,\,\theta,\,\phi\right) \;\bigg|\; v + 2 A S(\theta)-f(2m,\theta) = \lambda \bigg\},
\label{foliation}
\end{equation}
where $\lambda$ is the foliation label, $S(\theta)$ is the function defining the foliation and $A = 2m \rme^{-2u_0}$. The $2A$ factor and the 
extra $f(2m,\theta)$ are inserted to simplify future calculations. Note however that for $S(\theta)=0$, 
this transformation essentially reverses the Eddington-Finkelstein transformation (\ref{EFtrans}) at $r=2m$. 

We take the outward-pointing null normal to be the Killing vector $\ell = \partial/\partial v$: this is null vector field that will map 
leaves of the foliation (\ref{eq:surfaces}) into each other. 
The inward null normal will necessarily be a linear combination of 
$\ell_a = \rme^{2 U -2u_0} \left[ dr \right]_a$ and (the four-dimensional)
\begin{equation}
d \lambda = dv + \left( 2 A \frac{d S}{d \theta} -  \frac{\partial f}{\partial \theta} \right) d \theta \, . 
\end{equation}
%
Requiring that $\ell^an_a = -1$ and $n^a n_a = 0$ we find
\begin{equation}
\fl
n_a = - \left[ dv \right]_a + 2 \rme^{-2u_o} \left(\frac{d S}{d \theta}  \right)^2 \left[ dr \right]_a + 
\left(- 2A \frac{d S}{d \theta} + \frac{\partial f}{\partial \theta} \right)  \left[ d \theta\right]_a \, , 
\end{equation} 
with $f=f(2m,\theta)$. 
It is then a straightforward (though somewhat messy) calculation to check that the outward null expansion vanishes and the inward null expansion is 
\begin{eqnarray}
\fl
\theta_{(n)} =  -\frac{2\rme^{-2U}}{A} \left\{   \frac{d^2 S}{d \theta^2} 
- \left(\frac{d S }{d \theta} \right)^2 
+ \cot \theta  \left(\frac{d S }{d \theta} \right)
 -  2m \frac{\partial U}{\partial r} + m \frac{\partial V}{\partial r} + 1 \right\} 
 \label{tn}
\end{eqnarray}
where all quantities are evaluated at $r=2m$. 

We can eliminate the $r$ derivatives along with references to $V(r,\theta)$ by applying (\ref{E2}) and (\ref{E3}):
\begin{equation}
\fl
\theta_{(n)} =   -\frac{2\rme^{-2U}}{A} \left\{   \frac{d^2 S}{d \theta^2} 
- \left(\frac{d S }{d \theta} \right)^2 
+ \cot \theta  \left(\frac{d S }{d \theta} \right)
+ 2 \cot \theta \left( \frac{\partial U}{\partial \theta} \right)  - \left( \frac{\partial U}{\partial \theta  } \right)^2  + 1 \right\}  \, ,
 \label{tN}
\end{equation}
which, given (\ref{HorRel}), is significantly easier to work with. 
Note that
to ensure a smooth foliation and non-divergent $\tn$, the derivative of $S(\theta)$ must vanish at the poles. 
 
\subsection{An expression for $\delta_n \tl$}
Next, we show that if $\tn$ is everywhere negative then so is  $\delta_{\bar{n}} \theta_{(\bar{\ell})}$ for the scaling $\bar{\ell}$ determined
by the equivalence result (\ref{equivalence}). To begin we find
\begin{equation}
\tilde{\omega}_a =  \left(  \frac{\partial U}{\partial \theta} -   \frac{dS}{d \theta}  \right) [d \theta]_a  \, ,
\end{equation}
where here $U = U(2m, \theta)$. 
It is clear that $\omega_a$ is a total derivative. Recalling the discussion surrounding (\ref{AngMom}) this is perhaps not too much of a surprise 
since we are dealing with a static spacetime and so wouldn't expect the horizon to carry angular momentum. 
In any case, the form of the angular momentum one-form combines with $\ell = \frac{\partial}{\partial v}$ to suggest that 
\begin{equation}
 \bar{\ell} =  \rme^{2S-2U} \frac{\partial}{\partial v} \; \; \Rightarrow \; \;  \bar{\omega}_a =  \left(  \frac{dS}{d \theta}-   \frac{\partial U}{\partial \theta}   
 \right) [d \theta]_a\, . 
\end{equation}
is the desired scaling. 

To see this we calculate
\begin{equation}
\tilde{R} = \frac{\rme^{4 u_0 - 2 U}}{2m^2} \left( \frac{\partial^2 U}{\partial \theta^2} - 2 \left(\frac{\partial U}{\partial \theta}\right)^2 
+ 3 \cot \theta \frac{\partial U}{\partial \theta}  + 1 \right ) \, , 
\end{equation}
from which we find our final result by (\ref{dNtL}):
\begin{equation}
\fl
\delta_{\bar{n}}  \theta_{(\bar{\ell})} = -\frac{\rme^{-2U}}{ A^2} \left\{ \frac{d^2 S}{d \theta^2} - \left(\frac{d S}{d \theta}\right)^2 \! \! +\cot \theta \left(  \frac{d S}{d \theta} \right) +2  \cot \theta  \left( \frac{\partial U}{\partial \theta} \right) - \left( \frac{\partial U}{\partial \theta} \right)^2 + 1 \right\} \, . 
\label{dNtL_quad}
\end{equation}
Then, as expected, $\kappa \tn = \delta_{\bar{n}}  \theta_{(\bar{\ell})}$

\subsection{The determining equation}
Thus, the isolated horizon can only be foliated by future outer trapped surfaces if there exists a $g(\theta)$ satisfying both 
$g(0) = g (\pi ) = 0$ and 
\begin{equation}
G[g] = \frac{d g}{d \theta} - g^2 + g \cot \theta  + 2 U_\theta  \cot \theta - U_\theta^2 + 1 > 0 \, . 
\label{ME}
\end{equation}
We will refer to this inequality as the \emph{determining equation}.  Our investigation of horizons for this entire class of 
spacetimes reduces to an study of whether or not there exist $g(theta)$ that satisfy this inequality. 
That will be the subject of the next section. 

\section{Analysing the determining equation}
\label{sec:foliation}

\subsection{Examples}
\begin{figure}[h!]
	\begin{center}
		\subfloat[Quadrupole distortion]{\includegraphics[width=2.6in]{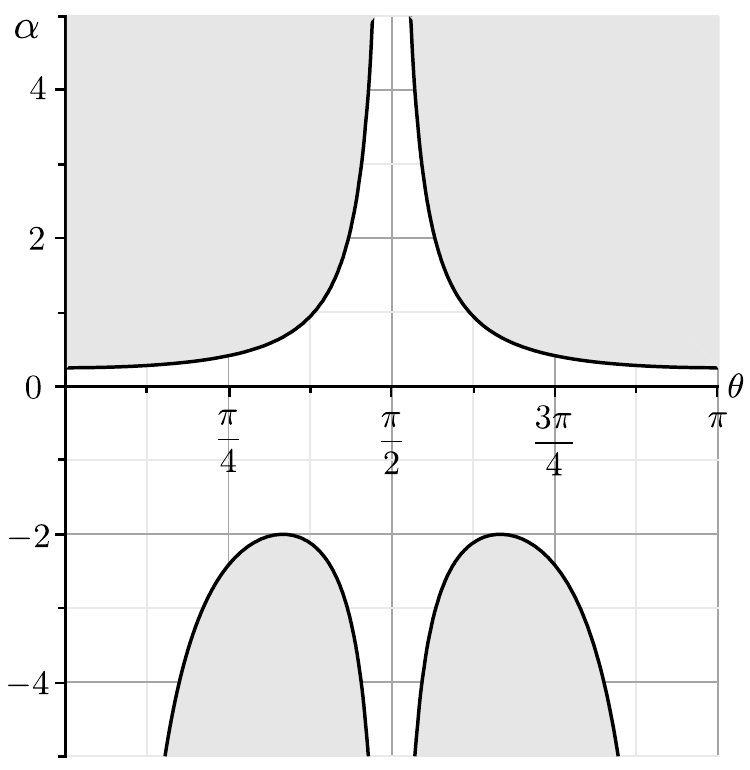}\label{Quad}}
		\subfloat[Dipole-octupole distortion]{\includegraphics[width=2.6in]{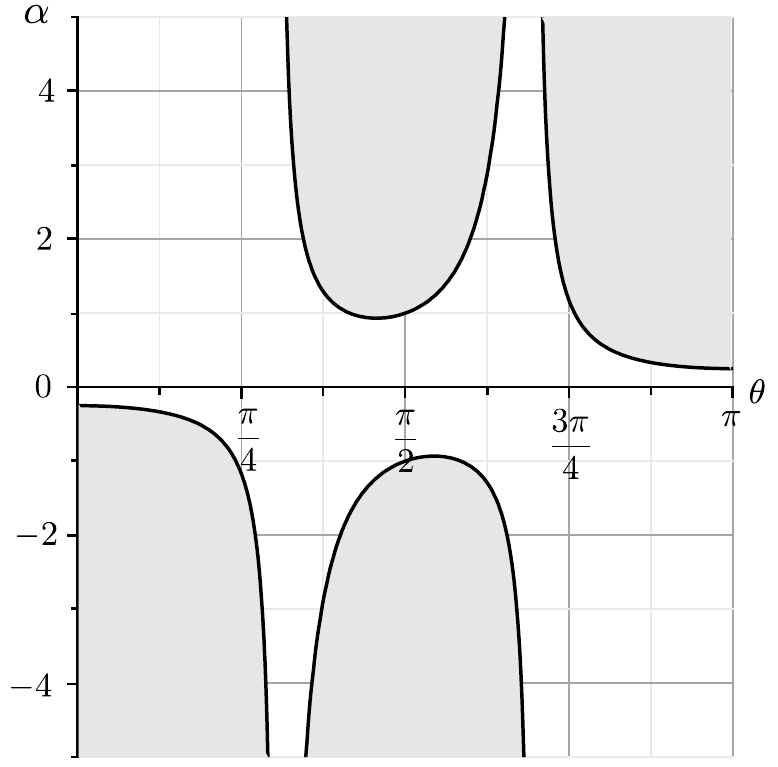}\label{DipOcto}}
	\end{center}
	\caption{Values of the distortion parameter for which $\tn < 0$ and $\delta_n \tl < 0$ when $g=0$. In the white areas they are negative
	while in the grey areas they are positive. }\label{G_Ex}
\end{figure}

To build intuition, we begin with a couple of special cases. We adopt the standard foliation (\ref{foliation}) with $S=0$ and 
examine the range of values of $\alpha$  for which the resulting leaves of the horizon foliation can immediately be seen to be FOTS. 
For the pure quadrupole case with $\alpha_2 = \alpha$:
\begin{equation}
G_{2}[0] = 1-  4 \alpha \cos^2 \theta - 4 \alpha^2 \sin^2 \theta \cos^2 \theta    \, .
\label{QuadEq}
\end{equation}
It is clear that this will not be everywhere positive and this is shown in Figure \ref{Quad}. For $-2 < \alpha < 1/4$, $G_2[0]$ is
positive but for more extreme distortions it becomes partially negative. 
%

Next consider the pure dipole-octupole case. Then by (\ref{eq:multipoles}) $\alpha_1 = - \alpha_3 = \alpha$ for some $\alpha$ and we have:
\begin{equation}
G_{13}[0] = 1 - 2 \alpha \cos \theta (1 - 3 \cos^2 \theta) - \alpha^2 \sin^2 \theta (1 - 3 \cos^2 \theta)^2  \, .
\label{DipOctEq}
\end{equation}
Again it is clear that this will not be everywhere positive and this is shown in Figure \ref{DipOcto}. For $-1/4 < \alpha < 1/4$, 
$G_{13}[0]$ is everywhere positive but for more extreme distortions it also takes negative values. 
In both cases 
increasing the distortion parameter not only decreases the range of positive values of $G[0]$ but also causes the 
negative regions to become increasingly negative. 

The classification problem for these isolated horizons then amounts to deciding whether or not there exists a function $g(\theta)$ such that
\begin{equation}
G[g]-G[0]= \frac{d g}{d \theta} - g^2 + g \cot \theta 
\label{CT}
\end{equation}
can be made to balance off the negative areas arising from $G[0]$ while still keeping the positive areas positive. This is a delicate balancing act.
 
\begin{figure}
\begin{center}
\includegraphics[width=2.6in]{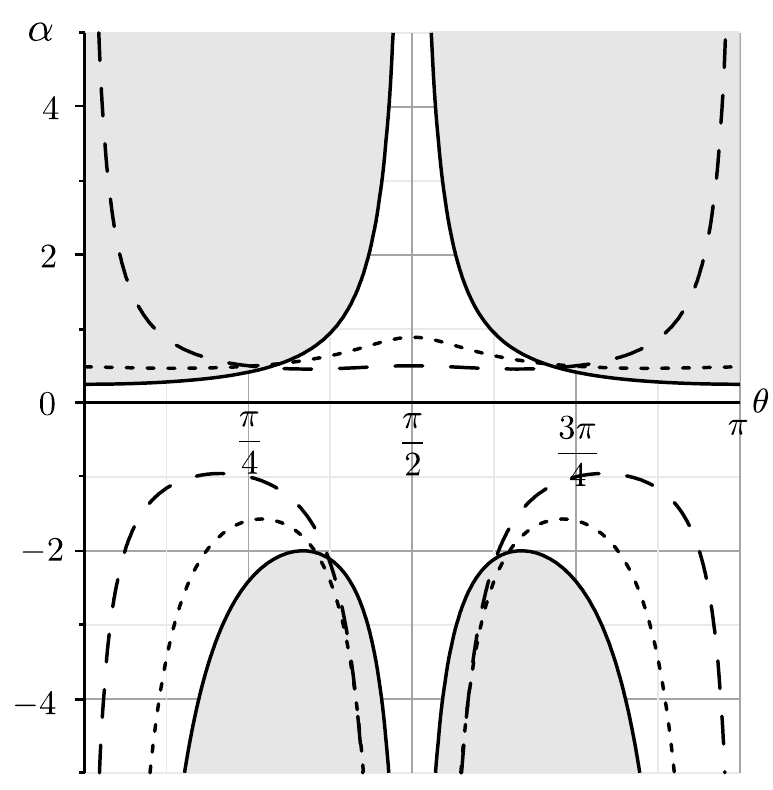}
\caption{The allowed ranges of distortions for $g_1$ (dashed) and $g_2$ (dotted). In both cases the maximum allowed (positive) $\alpha$
increases. The allowed range is that above/below the lines. }
\label{G_lm}
\end{center}
\end{figure}

For the quadrupole case, it is not too hard to (slightly) improve the upper bound on $\alpha$: large positive contributions can only be obtained from (\ref{CT}) by having $g$ rapidly decrease in the regions of interest. For example choosing
\begin{equation}
g_1 = \alpha \sin 2 \theta \; \; \mbox{or} \; \; g_2 = - \alpha \, \mbox{erf} (\theta-\pi/2) \sin \theta\, , 
\end{equation}
will increase the maximum distortion to about $\alpha \approx 0.45$ as shown in Figure \ref{G_lm}. At the same time it decreases the allowed
range of values for negative $\alpha$. This is okay as we would expect to have to pick different $g$ for different values of $\alpha$.  Note however
that this increase of maximum allowed distortion has come at a cost: the region around $\pi/2$ is no longer positive for all $\alpha$. In fact, increasing the factor of $\alpha$ in 
$g_1$ and $g_2$ to, for example, $2 \alpha$ will actually result in a decrease in the maximum allowed distortion as the the curve begins to dip towards
zero for $\pi/2$. 

A bit of experience trying to construct functions to increase the allowed range of distortions to larger (absolute) values of $\alpha$ quickly leads one
to suspect that these kinds of behaviours are generic: most choices of $g$ do not improve the range and those that do only increase it by 
fairly small amounts. In particular it seems very unlikely that one could construct a $g$ that would give $G[g]>0$ for arbitrarily large values of $\alpha$. 

\subsection{Large $\alpha$ quadrupole distortions}
We now substantiate this suspicion with the following theorem for quadrupole distortions which demonstrates that in general one cannot always
foliate distorted isolated horizons so that $\tn < 0$ everywhere. Similarly  it is not always possible to choose a scaling of the null vectors 
so that $\delta_n \tl < 0$ everywhere. 
\begin{theorem}
For $|\alpha| \geq 5$, there is no differentiable function $g(\theta)$ satisfying
\begin{equation}
\frac{dg}{d\theta} > g^2 - g \cot \theta - 1 + 4 \alpha \cos \theta^2 + 4 \alpha^2 \sin^2 \theta \cos^2 \theta 
\label{Keq}
\end{equation}
over the entire interval $0 \leq \theta \leq \pi$. 
\end{theorem}
Note, that there is no suggestion that $|\alpha| = 5$ is tight. As we are about to see, there are several crude assumptions
in the proof and so this bound can certainly be improved (for example, it is easy to show that $\alpha >3$ is sufficient for positive $\alpha$). 
However, in this paper we are mostly concerned with the principle and so
 won't worry about optimizing the bounds.

\textbf{Proof:} The basic idea is consider the possible values of $g(\pi/2)$. We show that if $g(\pi/2) \geq 0$ then $g$ necessarily diverges
somewhere between $\pi/2$ and $\pi$. On the other hand if $g(\pi/2) < 0$ then $g$ necessarily diverges between $0$ and $\pi/2$. Thus, there is 
no continuous $g$ that will everywhere satisfy (\ref{Keq}).  The root of the
divergence is the $dg/d\theta > g^2$ part of (\ref{Keq}). The rest of the proof is just working to ensure that there are no strongly negative
quantities around which might counter this divergence.  


Now, the details. We start by assuming that $g(\pi/2) \geq 0$. Then, on completing the square for $g$ in 
(\ref{Keq}), we obtain:
\begin{equation}
\frac{dg}{d\theta} > \left( g - \frac{1}{2} \cot \theta \right)^2 + Z \, ,
\label{Keqp}
\end{equation}
where
\begin{equation}
Z = -1 - \frac{1}{4} \cot^2 \theta + 4 \alpha \cos^2 \theta + 4 \alpha^2 \sin^2 \theta \cos^2 \theta \, . 
\end{equation}
For purposes of the proof, we wish to work over a range of $\theta$ where we can guarantee that:
i) $g(\theta)> g_o$ where $g_o$ is a reasonably large positive constant and ii) $g(\theta)$ is increasing. $g(\pi/2)$ is not a suitable choice for $g_o$;
while $g$ is non-negative by assumption,  $Z(\pi/2) = -1 < 0$ and so the derivative of $g$ could be negative. For
$|\alpha| \geq 5$ we can only guarantee $Z > 0$ over the range $0.537 \pi < \theta < 0.843\pi$. 
 
To get to a point where our conditions will be satisfied, we note that 
the first term of (\ref{Keqp}) is always positive and so we can bound $g$ at any point $\theta$ by:
\begin{equation}
g(\theta) - g(\pi/2) > \int_{\pi/2}^{\theta} Z d \vartheta \, . 
\end{equation}
Choosing $\theta=3\pi/4$ we find
\begin{equation}
g(3\pi/4) > g(\pi/2) + \left(\frac{\pi}{8}\right) \alpha^2 + \left(\frac{\pi}{2} - 1  \right) \alpha - \frac{3 \pi + 4}{16} \, 
\end{equation}
or since $g(\pi/2) \geq 0$ and $|\alpha|\geq 5$, $g(3\pi/4) > 6.12$ (the lower bound is set by the negative $\alpha$). 
For our purpose $6.12$ is sufficiently large and with $3\pi/4$ in the range $0.537 \pi < \theta < 0.843\pi$ we know that $Z(3\pi/4)>0$
and so $dg/\theta$ is also positive at that point. 

$Z$ will remain positive until $\theta = 0.843\pi$ so until that point we can drop it from the right-hand side of (\ref{Keq}) and still have:
\begin{equation}
\frac{dg}{d \theta} >  \left( g - \frac{1}{2} \cot \theta \right)^2 > g^2 \, , \label{Keq2}
\end{equation}
(since $\cot \theta$ takes negative values over this range). Now the related differential equation
\begin{equation}
\frac{dG}{d\theta} = G^2 \, , 
\end{equation}
with initial condition $G(3 \pi/4) = G_o$ has solution
\begin{equation}
G(\theta) = \frac{1}{(1/G_o) + (3 \pi/4 -\theta)}
\end{equation}
and so diverges at $3 \pi/4 + 1/G_o$. Then if we define $\gamma$ by $g = G + \gamma$, (\ref{Keq2}) implies
\begin{equation}
\frac{d \gamma}{d \theta} > 2 G \gamma + \gamma^2 \, , 
\end{equation}
and with all functions positive, $\gamma$ must also diverge by $3\pi/4 + 1/G_o$. Taking $G_o = 6.12$ (the lower bound we found ealier) 
this implies that $\gamma$ (and so $g$) will necessarily diverge before $\theta = 0.81 \pi$. \

Thus, if $|\alpha| \geq 5$ then there is no function $g$ satisfying both $g(\pi/2) > 0$ and (\ref{Keq}) over the entire range. Setting 
$\vartheta = \pi - \theta$ and $\tilde{g} = - g$ exactly the same reasoning will demonstrate that similarly there is no 
solution $g(\pi/2) < 0$. If $g$ cannot be either positive, negative, or zero at $\pi/2$ then it is clear that there is no $g$
satisfying (\ref{Keq}) everywhere. This establishes the desired result.

\section{Discussion}
\label{Discussion}
We have seen that for sufficiently distorted Schwarzschild solutions, the isolated horizon can be foliated with neither marginally trapped
nor outer trapping two-surfaces. Equivalently no slice of the horizon can be infinitesimally deformed to become fully trapped or 
even just outer trapped. By Hayward's terminology, the trapping horizon is neither future nor outer (or for that matter past or inner!). 
At the same time we have seen that these horizons bound a region that contains both a singularity and trapped surfaces. Thus, there is good 
reason to continue to identify the horizon as a black hole boundary even if the exact extent of the prison has not been determined. 

In considering these examples, it should be kept in mind that quadrupole distortions of $\alpha = \pm 5$ are very large. For $\alpha = 5$, 
$\tilde{R}$ ranges between $-2 \times 10^{8} /m^2$ and about $1 \times 10^{10}/m^2$ while for $\alpha = -5$, it is flattened so that its
range is from $-1 \times 10^{-5}/m^2$ to $3 \times 10^{-5}/m^2$. Though these are exact solutions their physical relevance is, to say the least, 
unclear. For $\alpha > 0$ one is stuck with non-standard infinities: it has been shown that inducing such distortions in standard spacetimes
via matter fields requires violations of the strong energy condition. For $\alpha < 0$, in principle it should be possible to induce the distortions
with regular matter, however one suspects that at best extreme stresses would be required and it is quite possible that physical or geometric
considerations might prevent the piecing together of the desired solutions. We will consider the feasibility of these constructions in a future
paper. 

So, while these solutions appear to provide a counter-example to the idea that all static black hole solutions will be FOTHS or OTHs, 
their extreme nature needs to be kept in mind. It is quite possible that despite the fact that no trapped surfaces ``hug'' the isolated horizon,
it may still turn out that the horizon bounds the prison: it is just that the trapped surfaces approach it point-by-point rather than all at once. 
Alternatively it is conceivable, though we think less likely, 
 that the solutions contain one or more other isolated horizons (we only examined those of constant $r$). 
Then it is possible that the prison is bound by such a horizon and that horizon is a FOTH.  
It would be interesting to determine the true extent of the prison and so properly understand this final point. 

\section*{Acknowledgements} The authors acknowledge Jos\'{e} Senovilla in correcting some
erroneous comments that we made in the Introduction of the published version of this paper. This research was supported by the Natural Sciences and Engineering Research Council of Canada. 

\vspace{1cm}

\bibliographystyle{utphys}
\bibliography{research_reading}
\end{document}